# Longitudinal acoustic properties of poly(lactic acid) and poly(lactic-co-glycolic acid)

N G Parker[1], M L Mather[2], S P Morgan[2] and M J W Povey[1]

[1] School of Food Science and Nutrition, University of Leeds, Leeds, UK
[2] Electrical Systems and Applied Optics Research Division, University of Nottingham, Nottingham, UK
n.g.parker@leeds.ac.uk

**Abstract**. Acoustics offers rich possibilities for characterising and monitoring the biopolymer structures being employed in the field of biomedical engineering. Here we explore the rudimentary acoustic properties of two common biodegradable polymers poly(lactic acid) and poly(lactic-co-glycolic acid). A pulse-echo technique is developed to reveal the bulk speed of sound, acoustic impedance and acoustic attenuation of small samples of the polymer across a pertinent temperature range of 0-70°C. The glass transition appears markedly as both a discontinuity in the first derivative of the speed of sound and a sharp increase in the acoustic attenuation. We further extend our analysis to consider the role of ethanol whose presence is observed to dramatically modify the acoustic properties and reduce the glass transition temperature of the polymers. Our results highlight the sensitivity of acoustic properties to a range of bulk properties, including visco-elasticity, molecular weight, co-polymer ratio, crystallinity and the presence of plasticizers.

## 1. Introduction

The field of tissue engineering revolves around the creation of biological substitutes to replace, repair or regenerate human tissue [1,2]. Central to this is the provision of a mock extracellular matrix, or 'scaffold', upon which to grow the artificial tissue. Two recent landmark implantations of tissue-engineered windpipes [3,4] serve to illustrate the major advances that have been made in this direction, notably in stem cell therapy and tissue cultivation. A central element of these experiments is the use of natural, donated tissue scaffolds upon which the replacement tissue is grown. In the pursuit of reduced donor requirements, a concurrent challenge is the formation and refinement of artificial tissue scaffolds. These structures must closely mimic the native extracellular matrix in its structure, mechanical properties, permeability and biocompatibility. Additionally, the scaffold is commonly required to degrade controllably and safely in parallel with the growth of its natural counterpart, and release growth and differentiation factors to aid cell cultivation.

Of much interest as underlying biomaterials are the synthetic polyesters poly(lactic acid) (PLA) and poly(lactic-co-glycolic acid) (PLGA) [5]. These materials are acclaimed for their biocompatible degradation into naturally occurring acids, at a rate which can be tuned through the molecular properties of the polymer. Like the extracellular matrix they attempt to replicate, tissue scaffolds comprise of a solid framework and interspersed voids, and may be two- or three-dimensional [6]. Accurate and robust characterisation of scaffolds is crucial during the current research and development phase, and ultimately for quality assurance of end products. However, satisfying this requirement is non-trivial due to the diverse properties needed to adequately specify the scaffolds. This is particularly true for 3D foam scaffolds, with recent results suggesting a need to specify both structural (e.g., pore size, shape and surface area) and functional (e.g., fluidics, elasticity and permeability) properties [7-10].

Acoustics offer rich and yet largely untapped potential for the characterization of biopolymers and their structures. It has well-known attributes of being non-destructive, non-invasive, economical and able to probe inside many opaque media. Of importance for characterising the bulk properties of polymers is its ability to probe both elastic and viscous effects through sound speed and attenuation measurements. For example, the longitudinal speed of sound, $v$, of an isotropic elastic solid with density $\rho$ is given by,

$$v = \left( \frac{K + 4G/3}{\rho} \right)^{1/2}, \qquad (1)$$

which serves to demonstrate the intrinsic sensitivity to mechanical material properties (here the bulk and shear moduli, $K$ and $G$). In polymeric systems, acoustics has been used to study not just mechanical moduli [11, 12] but also crystallinity/disorder [13] and molecular weight [14, 15]. For composite structures, such as polymer foam scaffolds, acoustics is capable of concurrent scrutiny of structural, mechanical and fluidic properties [16]. Indeed, such possibilities are being exploited for characterization of bone [17], which shares common structural properties to these scaffolds, and more recently for determining the stiffness of titanium [18] and hydrogel scaffolds [19]. Additionally, acoustics has been applied to monitor the foam formation processes *in situ* [20] and the degradation of biopolymers [21].

The exploitation of acoustics for characterizing such structures requires *a priori* knowledge of the acoustic properties of the underlying biopolymer. However, there is a paucity of experimental data on this topic. To date, only Wu *et al* [21] have studied the bulk acoustic properties of PLA and PLGA biopolymers. Motivated by this, we here determine the acoustic properties of PLA and PLGA biopolymers over a range of temperatures. Using a pulse-echo scheme, we employ time-of-flight and reflection analyses to determine the bulk speed of sound and acoustic impedance, respectively. Our experimental set-up allows small volumes (~30 mm$^3$) of polymer to be considered, while still retaining high precision. The temperature dependence of the sound speed and attenuation are studied and reveal clearly the glass transition region through the accompanying visco-elastic changes in the material. We additionally study the effect of immersion in ethanol, which is commonly used as a pre-wetting agent in tissue culture [22], and whose plasticizing effect is found to cause major changes to the acousto-mechanical properties of the polymers.

In section 2 we outline the materials and methods involved in this study, from the preparation of our polymer samples (section 2.1) to the acoustic platform used to perform our measurement (section 2.2). The time-of-flight and reflectometry approaches that we use to derive the fundamental acoustic properties, that is the speed of sound, acoustic impedance and attenuation, are derived in section 2.3. In section 3 we present and discuss our results, beginning with the polymer speed of sound (section 3.1) and acoustic impedance (section 3.2) taken at a fixed temperature of 25$^o$C. We then examine the behaviour of the sound speed (section 3.3) and attenuation (section 3.4) with temperature and examine the role of the glass transition on these properties. Further, we examine the effect that immersion in ethanol has on the acoustic properties (section 3.5). Finally in section 4 we summarise our findings.

**2. Materials and methods**

2.1. Sample preparation
We consider poly(DL-lactic acid) (PLA) and poly(lactic-co-glycolic acid) (PLGA) biopolymers. Our biopolymer samples are provided by the School of Pharmacy at the University of Nottingham as foamed tissue scaffolds generated via the supercritical $CO_2$ method [23, 24]. They are rigid, opaque and white, approximately 12 mm in diameter, 6 mm in height and 0.2g in mass. We derive homogeneous solid samples from these foams as follows. Each scaffold is ground into a coarse powder and placed in a cylindrical aluminium mould with diameter 4 mm and height 5 mm. The mould is placed on a hot plate and heated to 150°C for approximately 1 hour at room pressure. This temperature greatly exceeds the glass transition temperature of the samples (which is in the vicinity of 40-60°C), causing the polymer to soften and relax into the mould. Care is taken to remove bubbles. Finally, the polymer is allowed to cool to room temperature, forming a disc-shaped solid sample, 4 mm in diameter and 2-3 mm in height. The samples are transparent with a slight yellow tint.

We consider four biopolymer samples – one sample of PLGA with polymer ratio 85:15, and three PLA samples of different molecular weight. These samples, and their basic properties, are summarized in table 1, along with the shorthand notation we will subsequently use. The raw polymers are Lakeshore Biomaterials[TM] biopolymers (SurModics Pharmaceuticals, Alabama). The weight average molecular weight of the raw polymer was determined by nuclear magnetic resonance. The supercritical $CO_2$ foaming method avoids solvent contamination and is not cited to cause polymer scission, thus we expect that the foamed biopolymer has the same molecular properties as the raw biopolymer. Immersion of the polymer in water (which is performed during the acoustic measurements) can be expected to modify the polymer material since PLA and PLGA degrade in water via hydrolysis into lactic acid, and lactic and glycolic acids, respectively. The quoted degradation half-lives in water are around 5-6 months for PLGA and over 12 months for PLA [5].

2.2. Acoustic measurement platform

Our acoustic measurements are performed using a scanning acoustic platform which is illustrated schematically in figure 1 and discussed in detail elsewhere [25]. This platform supports pulse-echo operation, in which a single piezoelectric transducer acts as both emitter and receiver of sound waves. In this study, we are concerned with echoes that arise from the reflection of sound at the sample interfaces. Consider sound waves in a medium denoted I, at normal incidence to an interface of a second medium denoted II. The reflection coefficient, that is, the ratio of the reflected wave amplitude to the incident wave amplitude, is given by,

$$R = \frac{Z_I - Z_{II}}{Z_I + Z_{II}}. \quad (2)$$

The acoustic impedance $Z$ is given by $Z=\rho c$, where $c$ is the speed of sound within the material. Thus whenever a sound wave encounters a boundary between two different acoustic impedances, an echo is generated. Due to continuity requirements, the wave is also partially transmitted, with the amplitude of the transmitted wave (relative to the incident amplitude) given by $T=1-R$. The sound transducer (Panametrics V311) has a central frequency of 10 MHz, a -6 dB bandwidth of 5 MHz, and a focal distance of $F=54$ mm. Along with the aperture diameter of $D$=13 mm, we determine the semi-aperture angle of the lens to be $\theta=\sin^{-1}(D/2F)=7^{\circ}$, which classifies the transducer as weakly focussing. Based on an aqueous medium at $20^{\circ}$C (for which the speed of sound $c$ is approximately 1500 ms$^{-1}$) the sound beam has a wavelength of $\lambda=c/f=150$ μm and a pulse length of 300 μm. In the focal plane the beam diameter will be diffraction-limited to $F_r=1.22\ F\lambda/D = 300$μm, while in the axial direction the depth of focus is $F_z=8F^2c/(D^2f+2Fc)=1.7$cm. Note that the depth of focus is considerably larger than the sample thickness and so by placing the focal plane at the midpoint of the sample we can ensure that it remains in focus along its whole axial extent.

| Species | Shorthand notation | MW (kDa) | $\rho$ (g cm$^{-3}$) | $T_g$ (°C) | Degradation time (months) |
|---|---|---|---|---|---|
| poly(lactic-co-glycolic acid) 85:15 | PLGA | 53 | 1.25 | 45-52 | 5-6 |
| poly(DL-lactic acid) | PLA15 | 15 | 1.20 | 45-55 | 12-16 |
| poly(DL-lactic acid) | PLA24 | 24 | 1.20 | 45-55 | 12-16 |
| poly(DL-lactic acid) | PLA60 | 60 | 1.20 | 45-55 | 12-16 |

Table 1: Overview of the biopolymers examined in this study. The weight average molecular weight (MW) is determined via NMR while the density $\rho$, glass transition temperature $T_g$ and degradation time are the guidelines quoted by the manufacturer.

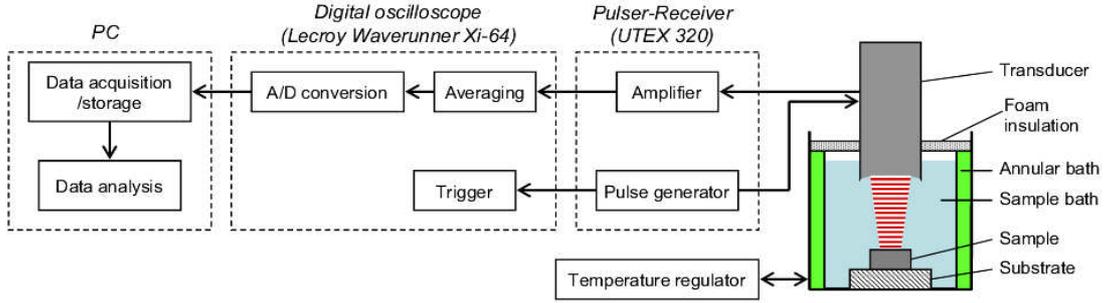

Figure 1: Block diagram of our acoustic measurement set-up.

A pulser-receiver (UTEX 320) provides a 300 V rectangular RF pulse of 10 ns duration to excite the piezoelectric transducer. This generates a sound pulse with a -6 dB duration of 150 ns, which corresponds to a spatial pulse length of approximately 250 μm in water. The pulse repetition rate is set to 500 Hz. Acoustic reflections returning to the transducer generate RF pulses which are received and amplified by the pulser-receiver. This signal is averaged over 200 sweeps by a digital oscilloscope (Lecroy Waverunner Xi-64) to reduce noise. Following analog-to-digital conversion the signal is exported to a PC for storage and further analysis.

The sample is insonified from above using water as a coupling fluid to convey the sound. Both the sample and coupling fluid are contained by a cylindrical aluminium sample bath. An outer annular container integrated into the sample bath acts as a temperature jacket. An aqueous solution of anti-freeze fluid is circulated through this jacket by a temperature bath/circulator (Haake B5/DC50). This enables us to regulate the temperature of the sample bath over the range -5 to $80^{\circ}$C. Independent measurements are taken of the coupling fluid adjacent to the sample using a temperature probe (Hart Scientific 5612 probe) with an accuracy of 0.01 $^{\circ}$C. Foam insulation restricts heat leakage from the bath.

2.3. Acoustic measurements

 2.3.1. Time-of-flight analysis
We instigate a time-of-flight analysis to determine the speed of sound in the polymer sample. Consider a homogeneous sample placed upon a substrate, immersed in a coupling fluid and insonified at normal incidence from above. From analyzing the time-of-flight of the three sound echoes shown in figure 2(a) we can isolate the speed of sound in the sample. In the absence of the sample, path A corresponds to reflection at the fluid-substrate interface. In the presence of the sample, path B corresponds to reflection at the fluid-sample interface and path C corresponds to reflection at the sample-substrate interface (after transmitting through the sample). We denote the speeds of sound in the fluid and sample by $c_1$ and $c_2$, respectively, and the transducer-substrate distance and sample thickness by $d_1$ and $d_2$, respectively. The times-of-flight for each path are given by,

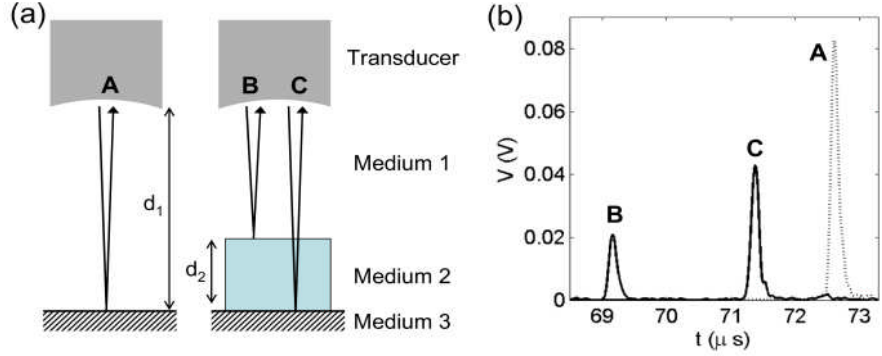

Figure 2: (a) Schematic of the time-of-flight measurements and (b) a typical received voltage. In both plots the three sound paths of interest, denoted A, B and C, are indicated. The trace in (b) corresponds to the PLGA sample immersed in water at 25°C.

$$t_A = 2d_1/c_1 \tag{3}$$
$$t_B = 2(d_1 - d_2)/c_1 \tag{4}$$
$$t_C = 2(d_1 - d_2)/c_1 + 2d_2/c_2. \tag{5}$$

Substituting equations (1) and (2) into equation (3) and rearranging gives,

$$c_2 = c_1\left(\frac{t_A - t_B}{t_C - t_B}\right), \tag{6}$$

from which we can readily determine the speed of sound of the sample, providing that of the reference fluid is known. A glass microscope slide acts as the substrate, with the large glass-polymer acoustic mismatch promoting vivid reflections. The voltage signal is demodulated computationally via the Hilbert transform and the pulse peak used to define its time-of-flight.

We have assumed that the sound waves travel perpendicularly to the fluid-sample interface and thus that the sonic path length through the sample is equal to its thickness $d_2$. In reality some sound rays deviate from normal incidence by up to the semi-aperture angle of the lens $\theta$. These extremal rays will undergo a path length that is larger than $d_2$ by a factor of $\sec\theta-1 \approx 0.007$. This is sufficiently small for this effect to be neglected in the current study.

When the sample is placed at the sonic focus, the diameter of the focal region (~300μm) is well within the diameter of the sample. Furthermore, the axial size of the sample (2-3 mm) is much greater than the acoustic wavelength (~150μm). Due to these dimensional criteria, we can expect that our acoustic measurements do indeed probe the bulk properties of the sample. The error in our measurement of the polymer speed of sound is dominated by the systematic error in locating the peak of each sound pulse which does not exceed 1%.

We employ both water (Millipore 0.22 μm) and ethanol as coupling fluids. The samples are immersed in the coupling fluid for 24 hours prior to analysis. The speed of sound of pure water has been accurately established over temperature and is determined from the fits of Bilaniuk and Wong [26, 27] (148-point), while the speed of sound of ethanol is determined using the fifth-order polynomial fit stated in [28]. These biopolymers are hydrophobic, suppressing the diffusion of water into the bulk. We therefore assume that, under immersion in water, the sample represents the true bulk of the pure polymer. The same is not true for ethanol, which is known for its affinity to these biopolymers [22].

Measurements are first taken at a fixed temperature of 25°C. After reaching the operating temperature we wait one hour for thermal stabilization and equilibration before taking measurements of pulses B and C. The sample is then removed so as to support pulse A. Following this we studying the temperature range -5–70 °C. The sample bath is first raised to the maximum temperature, and given time to equilibrate. Measurements are then taken during the cooling stage. The rate of cooling varies within the range of 30-60 °C per hour.

*2.3.2. Reflectometry analysis*

By considering the reflection amplitudes of sound pulses A and B we can additionally and independently determine the acoustic impedance of the sample. Since the wave amplitude is sensitive to diffraction, such reflectometry techniques are prone to larger errors than the time-of-flight method. However, it offers convenience for characterisation and online monitoring purposes since it requires surface reflections only. Consider a sound pulse at normal incidence upon a flat surface at a distance $d$ from the transducer. We can express the spectral voltage of the reflected sound pulse as,

$$V(d,f) = H(f) R M(d,f) \exp[-2\alpha(f)d] V_0(f). \tag{7}$$

Here $H(f)$ represents the combined electromechanical transfer function for emitting and receiving the sound pulse, $R$ is the reflection coefficient of the fluid-surface, $M(d,f)$ describes the modulation of the voltage with distance due to beam focusing and spreading, and $V_0(f)$ is the amplitude of the excitation pulse. The exponential term models beam attenuation with attenuation coefficient $\alpha(f)$, and the factor of 2 arises since the sound travels twice the transducer-surface separation $d$. Returning to the experimental set-up of figure 2(a), we can form an expression for the voltage emanating from the fluid-substrate interface, $V_A$, by substituting $d=d_1$ and $R=R_{13}$ into equation (7). Similarly, the voltage emanating from the fluid-sample interface, $V_B$, is given by inserting $d=d_1-d_2$ and $R=R_{12}$ into equation (6). Taking the ratio of $V_B$ to $V_A$, we can rearrange to give the following expression for the reflection coefficient of the water-sample interface,

$$R_{12} = R_{13} \frac{V_B(f)}{V_A(f)} \frac{M(d_1,f)}{M(d_1-d_2,f)} \exp[-2\alpha(f)d_2]. \tag{8}$$

That is, the reflection coefficient of the fluid-sample interface $R_{12}$ can be determined by reference to that of the fluid-substrate interface $R_{13}$. Once $R_{12}$ is determined we can then trivially evaluate the acoustic impedance of the sample by rearranging equation (2) to give,

$$Z_2 = Z_1 \left( \frac{1-R_{12}}{1+R_{12}} \right), \tag{9}$$

where $Z_1$ is the acoustic impedance of the coupling fluid. NB If the experiment was modified such that the sample and substrate interfaces were at the same position, then the beam modulation function would drop out completely. However, for convenience we allow for these interfaces to be at different vertical positions. To optimise the signal-to-noise ratio of the voltage measurements it is beneficial that the voltage signals from pulses A and B are of similar amplitude. For this reason we here change the substrate from glass to polyether ether ketone (PEEK), whose acoustics properties [29] can be expected to be closer to that of the biopolymers. From the determined values of speed of sound and acoustic impedance we can further determine the material density from the relation $\rho=Z/c$.

We determine all of the terms on the right-hand side of equation (8) as follows. $V_B(f)$ and $V_A(f)$ are obtained by Fourier transforming the received voltage pulses, while the distances $d_1$ and $d_2$ are obtained from the time-of-flight analysis. The attenuation and acoustic impedance of water are determined from [30], the acoustic impedance of PEEK is taken from [29], and equation (2) is used to determine $R_{13}$. Finally, the beam modulation function is to be calibrated experimentally under the following methodology. At a given frequency, the received voltage will reach a maximum value $V_{max}$ at the focal distance. At this maximum point we define $M(d,f)$ to be unity, such that equation (7) reduces to,

$$V_{max}(f) = V(d=F,f) = H(f) R(f) \exp[-2\alpha(f)F] V_0(f), \tag{10}$$

Dividing equation (7) by equation (10), and rearranging for $M(d,f)$, we arrive at,

$$M(d,f) = \frac{V(d,f)}{V_{max}(f)} \exp[2\alpha(f)(d-F)]. \tag{11}$$

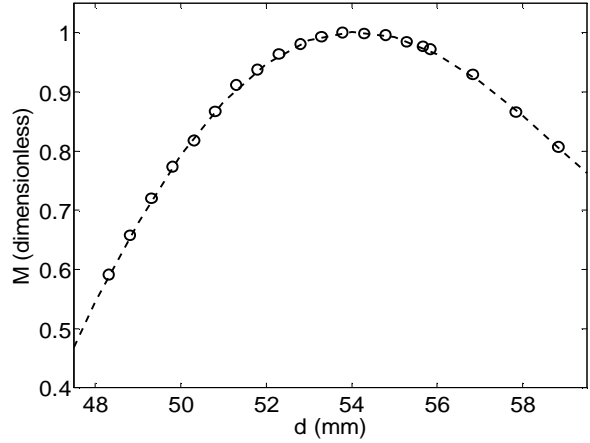

Figure 3: Beam modulation function $M(d, f=10$ MHz) determined from a PEEK substrate in water at 25°C. A fourth-degree polynomial ($M_{fit} = \sum_{n=0}^{n=4} a_n d^n$ with $a_0=277.987$, $a_1=-2.283\times10^4$, $a_2=6.867\times10^5$, $a_3=-8.982\times10^6$ and $a_4=4.325\times10^7$) provides an excellent fit ($R^2=0.9995$). We assume $\alpha=2.113\times10^{-14}$ Np m$^{-1}$ Hz$^{-2}$ [30].

These equations are not just a function of position and frequency, but also of temperature. Due to this large parameter space we restrict our impedance analysis to a single temperature (25°C) and frequency (the centre frequency of the transducer). The received voltage is Fourier transformed to reveal this frequency component, and the positioning system of the acoustic platform is used to translate the transducer over a range of $d$. The resulting beam modulation function is presented in figure 3. This function is independent of the substrate material, under the assumptions that the substrate is smooth, thick (free from resonance effects) and normal to the incident beam. As expected the peak occurs at the focal point $F=54$ mm, at which point the rays are in phase and the beam is at its narrowest. The function is smoothly varying and well behaved over the distance considered. An excellent fit to the data is achieved with a fourth-degree polynomial (see caption for polynomial specifications), to be used in the subsequent calculations of impedance.

*2.3.3. Attenuation analysis*

While the speed of sound probes the elastic properties of the material, one can also probe the viscous properties by considering the attenuation of sound as it traverses through the polymer. We will estimate the acoustic attenuation of the polymer by further examination of the received signal from the insonified sample and substrate. From equation (7) $V_B$ is,

$$V_B(f) = H(f) R_{12} M(d_1 - d_2, f) \exp[-2\alpha_1(f)(d_1 - d_2)] V_0(f). \quad (12)$$

Pulse C additionally transmits through the fluid-sample interface, is attenuated by the polymer, reflects from the sample-substrate interface, and transmits back through the fluid-sample interface. $V_C$ is then expressed as,

$$V_C(f) = H(f)(1 - R_{12})^2 R_{13} M(d_1, f) \exp[-2\alpha_1(f)(d_1 - d_2) - 2\alpha_2(f) d_2] V_0(f). \quad (13)$$

Close to the focal region the beam modulation function varies by less than 5% over the size of the samples (2-3 mm). Since the attenuation changes are considerably greater than this we will neglect the variation of $M$. Dividing $V_B$ by $V_C$ and rearranging for the polymer attenuation $\alpha_2$ we arrive at the expression,

$$\alpha_2(f) = \frac{1}{2d_2} \ln\left[ \frac{V_{12}(f) R_{13} (1 - R_{12})^2}{V_{13}(f) R_{12}} \right] \quad (14)$$

To enhance the amplitude of the transmitted pulse we employ the glass substrate. The reflection coefficients $R_{12}$ and $R_{13}$ are calculated from equation (2). We determine the acoustic impedance of water and the polymers from the relation $Z=\rho c$, using the temperature-dependent speeds of sound and fixed values for density. This density approximation is justified for liquids and solids since the temperature variation of density is usually much smaller than that of the speed of sound [15] (e.g., over the range 0-60°C the speed of sound in water varies by 10% while its density varies by less than 1%). To further support this approximation, it has been found that density changes play a minor role in the speed of sound variation in polystyrene [15], which bears strong physical similarities to PLA [31]. The acoustic impedance of the glass substrate is taken to be $Z=13.4$ MRayl. While we expect that the errors in our attenuation measurements may be as large as 10%, the large variations of attenuation permit at least qualitative interpretation of the data.

3. Results and discussion

We begin by analysing the speed of sound and acoustic impedance of the polymers at a fixed temperature of 25°C in sections 3.1 and 3.2 respectively. In section 3.3 we consider the effect of temperature on the speed of sound, followed in section 3.4 by the polymer attention. Finally, in section 3.5 we consider the effect of ethanol on the speed of sound.

3.1. Speed of sound (25 °C)

Using the time-of-flight method (section 2.3.1) we determine the bulk speed of sound in the polymers at a fixed temperature of 25°C with water as the coupling fluid. A typical voltage signal is shown in figure 2(b), with the three echoes clearly evident, and the measured data is presented in table 2. The PLA samples have a speed of sound of around 2260 ms$^{-1}$ while that of the PLGA is higher at 2326 ms$^{-1}$. To our knowledge the only similar study was by Wu *et al.* [21], who observed the speed of sound in un-degraded PLA to be around 2300 ms$^{-1}$ at 20°C, in close agreement with our findings. They also observed the speed of sound of PLGA 50:50 (compression moulded) to be 2450 ms$^{-1}$, somewhat larger than our PLGA sample. We attribute this to the different glycolic acid content. Indeed, Wu *et al.* measured the speed of sound of poly(glycolic acid) to be around 3000 ms$^{-1}$, compared to 2300 ms$^{-1}$ for PLA, indicating that the speed of sound increases with glycolic acid content.

| Sample | Molecular weight (kDa) | Time-of-flight | | | | Reflectometry | | Glass transition |
| --- | --- | --- | --- | --- | --- | --- | --- | --- |
| | | $Z_2$ (MRayl) | $t_A$-$t_B$ (μs) | $t_C$-$t_B$ (μs) | $c_2$ (ms$^{-1}$) | $Z_2$ (MRayl) | Density (kg m$^{-3}$) | $T_g$ (°C) |
| PLGA | 53 | 2.76 | 3.45 | 2.22 | 2326.0 | 2.76 | 1190 | 31.5 |
| PLA15 | 15 | 2.57 | 1.43 | 0.95 | 2252.9 | 2.57 | 1140 | 32.1 |
| PLA24 | 24 | 2.56 | 1.30 | 0.86 | 2262.5 | 2.56 | 1130 | 46.6 |
| PLA60 | 60 | 2.77 | 2.40 | 1.58 | 2273.5 | 2.77 | 1220 | 32.9 |

Table 2: Results of our time-of-flight and reflectometry meausurements in water at 25$^0$C, (assuming $c_1$=1496.70 ms$^{-1}$). The glass transition temperature is also presented.

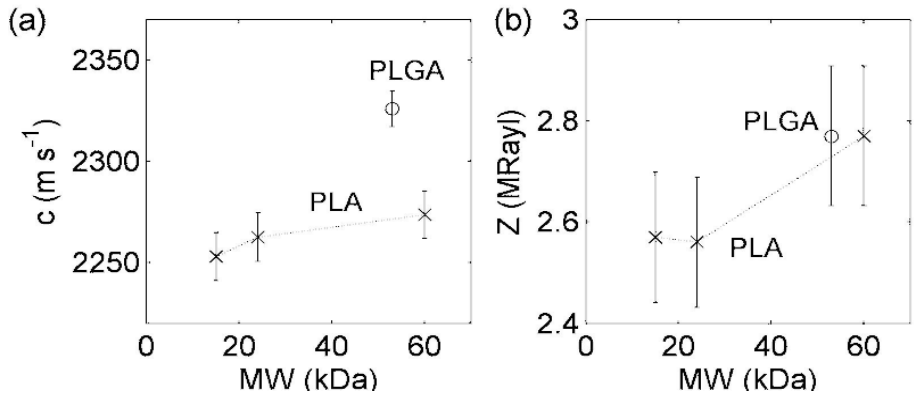

Figure 4: (a) Speed of sound $c$ and (b) acoustic impedance $Z$ of the biopolymers versus their weight average molecular weight MW, at an ambient temperature of 25°C.

In figure 4(a) we plot the polymer speed of sound versus molecular weight. For PLA we observe that the sound speed increases with molecular weight. A consistent observation was made in [21] (albeit for PLGA) where the degradation-induced reduction in molecular weight led to a reduced sound speed. The amount of free volume between molecules is known to increase with molecular weight, leading to enhanced cross-linking and steric effects, which will increase the rigidity of the polymer. Indeed, the elastic modulus of PLA has been observed to increase with molecular weight, varying sharply at low weights and tending to saturate at high weights [32]. We attribute our observed increase in sound speed with molecular weight to be dominated by this behaviour of the elastic modulus. A more detailed study of the sound speed with molecular weight could reveal whether the wider elastic moduli behaviour is followed. Furthermore, this could also promote a simplified model for sound speed in these polymers. For example, for liquid polymers the sound speed closely follows a simple formula based on the number of end groups and repeat units, their effective compressibilities and partial molar volumes [33], and tends to increases with molecular weight. This may share a common physical basis with the solid polymers considered here. However such a study is beyond the scope of this current work.

3.2. Acoustic impedance (25°C)

We determine the acoustic impedance of the polymers via the reflectometry method (section 2.3.2) encapsulated by equations (8) and (9). Working at the centre frequency of 10 MHz and a temperature of 25°C, the evaluated acoustic impedances are presented in table 2. While this method is subject to relatively large errors compared to the time-of-flight approach, it provides an independent corroborative measure. The values lie in the region ~2.6-2.8. Using our measured speeds of sound and acoustic impedances we can estimate the polymer density via $\rho=Z/c$, with the resulting values also presented in table 2. The values lie in the range $\rho$=1.13-1.21 g cm$^{-3}$, which is consistent with the manufacturer estimate of $\rho$=1.2 g cm$^{-3}$. We observe a slight increase in the density with the molecular weight. The variation of the acoustic impedance with polymer weight (figure 4(b)) indicates a similar trend to that for the speed of sound, increasing with molecular weight.

3.3. Temperature dependence of the speed of sound

The variation of the biopolymer speed of sound with temperature is presented in figure 5. All four samples show the same qualitative behaviour. The speed of sound decreases with temperature throughout, with distinct linear dependences in the low and high temperature regimes. This trend has been observed in the speed of sound-temperature curves for polystyrene [12, 14]. The low temperature regime corresponds to the hard, brittle glass phase of the polymer, while the high temperature regime is attributed to the flexible, soft rubber phase of the polymer. The smooth variation between the two linear regimes corresponds to the glass-rubber phase transition. The glass-rubber phase transition is a second order phase transition. Such transitions are characterized by a discontinuity in the first derivative of specific volume with temperature. Acoustics is sensitive to the specific volume, since the material density (the inverse of specific density) is a key parameter is determining the speed of sound (see equation (1)). Furthermore, acoustics has an additional sensitivity to a material's elastic modulus, which is further transformed during a phase change. Thus the speed of sound provides a useful system parameter with which to monitor phase changes. Indeed, the sharp change in $dc/dT$ present in figure 5 is strong evidence for this second order phase transition. It is interesting to note that we detect no significant variation in the polymer density with temperature and so, according to equation (1), the modulation of the speed of sound must arise dominantly from the modified elastic moduli. A similar conclusion is made regarding the sound speed in polystyrene [15], which is known to have similar thermo-physical properties to PLA [31]. The glass transition temperature is not constant but depends on the timescale of the experiment and thus on the sound frequency. However, for a similar polymer it has been shown that the acoustic measurements are insensitive to frequency across a broad range extending from 500 kHz to 500 GHz range [12, 14] due to the absence of significant structural relaxations therein. By analogy we can expect that our measurements at 10 MHz lie well within this frequency-insensitive regime.

The glass transition temperature can be estimated from the curves in figure 5 by fitting the rubber and glass regimes with straight lines and defining $T_g$ to be their crossing point [12, 14]. The predicted values for $T_g$ are included in table 2. For PLGA, PLA15 and PLA60, $T_g$ is around 32°C. This is considerably lower than the manufacturer's quote of 45-55°C, although PLGA and PLA samples with glass transition temperatures down to 27°C have been measured elsewhere [32, 34]. Several possibilities for this discrepancy are now suggested and discussed. The acoustic method we employ is distinct from the conventional approaches to determine the glass transition, e.g. differential scanning calorimetry, and so some deviations may arise from the different timescales and physical properties probed. While an exhaustive comparison between acoustic and other methods has not been made results indicate that the predictions should be in very close agreement [14]. While it is conceivable that water may enter the bulk and act as a plasticizer, the differing hydrophobicity between PLA and PLGA [35] should lead to significant deviations in their transition points, and no such variation is observed. Instead, we believe that the reduced transition temperature is due to the degradation of the polymers during the course of this study. Such degradation will reduce the average molecular weight of the polymer, which is cited to lower the glass transition in polystyrene [14]. Thermally-induced degradation may occur during sample preparation through polymer scission and unzipping events [36, 37]. However, Park *et al.* examined the glass transition temperature after various heating cycles in an inert nitrogen atmosphere and observed no decrease in $T_g$. The likely candidate is thus hydrolysis, which may occur during both the sample preparation and the immersion measurements. While the timescale of the hydrolysis at room temperature is expected to be of the order of months, the raised immersion temperatures here may dramatically accelerate this degradation. Houchin *et al* showed that the hydrolysis of PLGA was dramatically accelerated by raised temperature and ambient water [38]. Indeed, following heated immersion we find that the samples develop a thin milky residue on their surface likely to be the hydrolytic products, while no such residue is observed when the samples are immersed at room temperature. In anomaly, the glass transition for PLA24 is evaluated to be around 46 °C. Given that the same experimental protocol was performed on every sample, we attribute this to anomalous properties in the origin polymer sample: polymer degradation via hydrolysis has been shown to be strongly dependent on environment, composition and history [39].

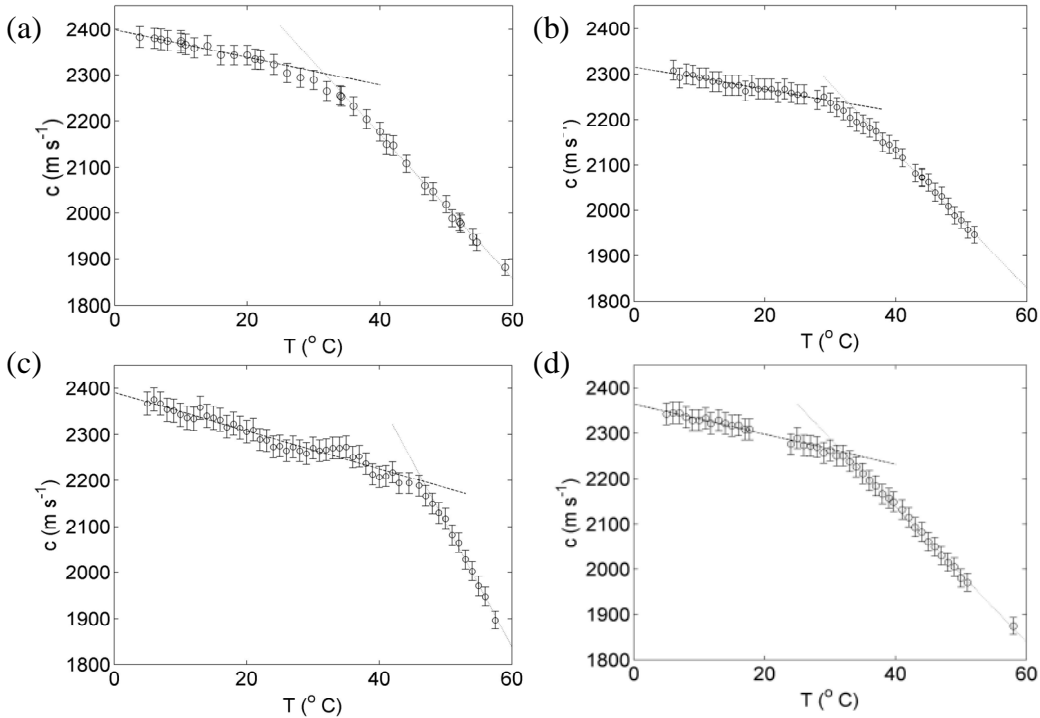

Figure 5: Biopolymer speeds of sound as a function of temperature for (a) PLGA, (b) PLA15, (c) PLA24 and (d) PLA60. In the upper and lower temperature regimes straight lines have been fit to the data.

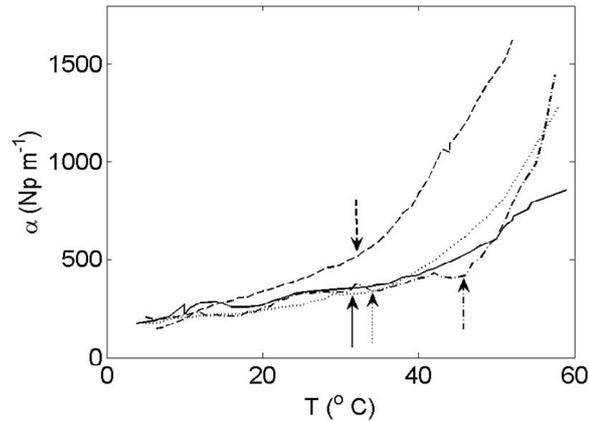

Figure 6: Acoustic attenuation versus temperature for PLGA (solid line), PLA15 (dashed line), PLA24 (dot-dashed line), PLA60 (dotted line). Arrows indicate the glass transition temperature for each polymer, as determined from speed of sound measurements.

3.4. Attenuation
The polymer attenuation, calculated from equation (14), is plotted in figure 6 as a function of temperature. In the glass phase the attenuation increases with temperature. Upon transiting to the glass phase the attenuation increases much more rapidly. This is most evident for the PLA15 sample (dashed line). Consider the material to be wholly amorphous. Below the glass transition the molecules are in a disordered arrangement that is locked in place to form a metastable state. Due to the rigidity of the structure a sound wave will propagate through in a highly elastic manner. However, at the glass transition, the free volume within the structure becomes sufficiently large to allow segments of the polymer chain to undergo motion while the ends of the molecule remain firmly pinned in place, primarily in the form of 'crank shaft' rotations. The passing sound wave will excite these modes, causing inelastic loss of energy from the wave. Thus the material behaves in a highly viscous manner in this region. Indeed, the large attenuation of the transmitted sound pulses ultimately limits the uppermost temperature we can probe in our experiments. We also note that a sharp increase in sound attenuation at the glass transition has been observed in polystyrene [12]. It is important to note that these effects apply to the amorphous components of the material only. In reality we can expect the presence of some regions of polymer that are in the lowest energy crystalline configuration and will propagate sound in a highly elastic manner. Thus the acoustic attenuation above $T_g$ should provide a particularly sensitive probe of the degree of crystallinity in the material. Indeed, a higher fraction of amorphous solid may explain the anomalously large attenuation in the PLA15 sample.

3.5. Effect of ethanol on the speed of sound

In figure 7 we present the speed of sound-temperature curves following immersion in ethanol for 24 hours. Note that a further 24 hours immersion had no significant effect on the behaviour. Ethanol acts as a plasticizer on the polymer and has a marked effect on its acoustic properties. For the PLGA sample the speed of sound is reduced by approximately 10%, while for the PLA samples the shift is considerably larger and up to 40%. This difference in ethanol sensitivity is most likely due to differences in the crystallinity between the PLA and PLGA samples. The samples feature amorphous and crystalline regions with only the former undergoing plasticization. Thus the weaker effect observed in the PLGA sample suggests a larger crystalline component to the solid. (Differences due to the polymer-ethanol interaction can assumed to be negligible since PLA and PLGA have similar affinities for ethanol [40].) We observe that the perfusion of ethanol results in a decrease of the polymer density by around 10% across all of the samples. According to equation (1), a reduction in density will tend to increase the sound speed. Thus the observed decrease in sound speed suggests that the elastic modulus is greatly reduced by the presence of ethanol.

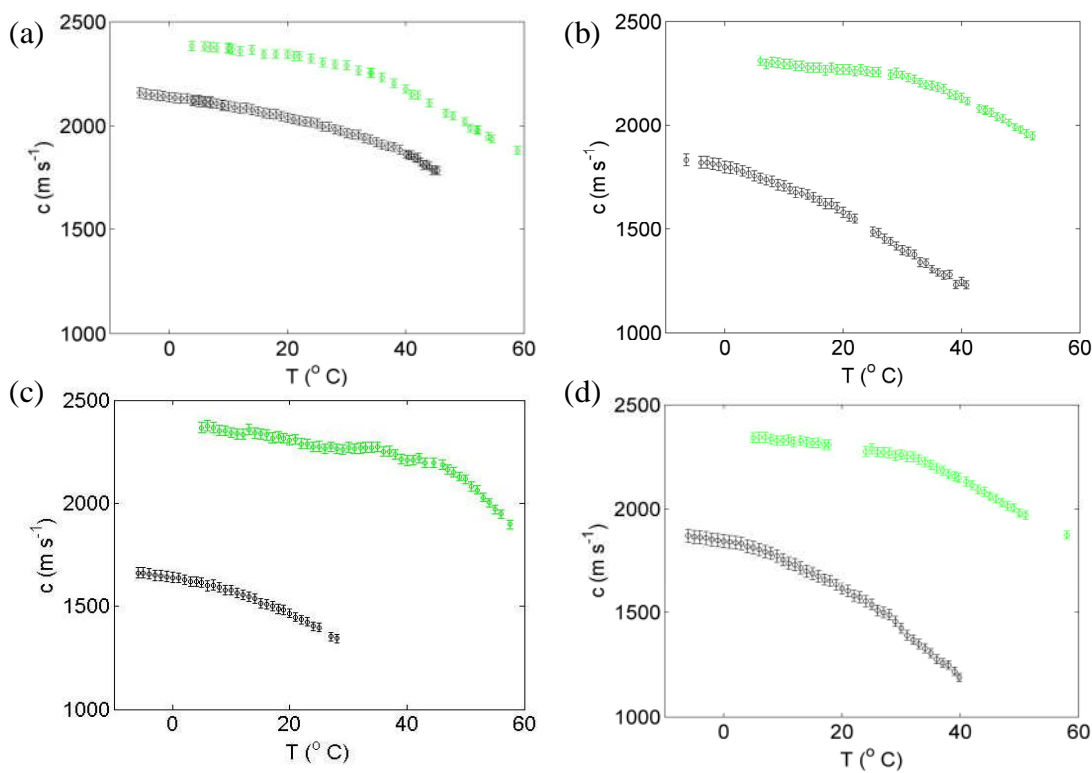

Figure 7: Speed of sound of the biopolymers under immersion in ethanol (black data points). Shown are (a) PLGA, (b) PLA15, (c) PLA24 and (d) PLA60. For comparison, the speeds of sound under immersion in water are shown in green/grey.

We observe that the glass transition temperature becomes reduced in the presence of ethanol. This is most clearly evident for PLA15 and PLA60, with the glass transition now occurring at approximately 5$^{o}$C. Ethanol acts as a plasticizer, increasing the free volume within the polymer and the separation between polymer molecules, and reducing inter-chain interactions. Its plasticizing effects on PLGA have been observed by Ahmed *et al*. who found the glass temperature to reduce from 44$^{o}$C to 11$^{o}$C following incubation in ethanol [41]. The glass transition also becomes broadened, most notably for PLGA and PLA24. Broadening of the glass transition during plasticization suggests that the plasticized state is heterogeneous, and this may be encouraged by the presence of crystalline domains within the sample [42].

**4. Conclusions**

In this work we have analysed the longitudinal acoustic properties of poly(lactic acid) and poly(lactic-co-glycolic acid). A simple pulse-echo experimental geometry enabled the acoustic properties to be determined using small sample volumes (of the order of 30 mm$^3$) and over a large temperature range (-5 - 70$^{o}$C). By analysing the acoustic time-of-flight and reflection amplitudes we determine the speed of sound, acoustic impedance and attenuation of the samples. At 25$^{o}$C the speed of sound lies in the vicinity of 2200-2300 ms$^{-1}$ for all of the samples considered. The variation amongst samples suggests that the speed of sound increases with both the molecular weight and the co-polymer ratio of glycolic acid.

The temperature dependence of the speed of sound reveals clearly the presence of a second-order phase transition that corresponds to the glass transition of the polymer. The detected glass transitions lie in the range 32-46$^{o}$C. Furthermore, upon passing from the glass to the rubber phase the acoustic attenuation increases sharply, and is attributed to the inelastic restructuring of the 'unlocked' polymer molecules by the passing compression wave.

In tissue engineering, ethanol is used as a pre-wetting agent to enhance cell culturing and adhesion onto tissue scaffolds composed on these biopolymers. Based on this we extended our analysis to consider ethanol as the ambient fluid. We observe significant reductions in the glass transition temperature to values of around 5$^{o}$C. An associated broadening of the glass transition restricts an accurate estimation of the transition temperature. Furthermore, the perfusion of ethanol through the solid sample causes a decrease in the speed of sound, which can be related to a large-scale reduction in the mechanical stiffness of the polymer. These observations emphasize a risk in destroying the structural integrity of scaffolds during immersion in ethanol if its plasticizing effects are not accounted for.

Our observations highlight the acoustic sensitivity to molecular weight, co-polymer ratio, crystallinity and fluid perfusion and suggest routes to monitor these important quantities, subject to further investigation and characterisation. Moreover, the increased knowledge of the basic acoustic properties of these biopolymers may aid in the acoustic characterisation of composite biopolymer structures. Indeed, a recent demonstration of acoustic propagation through a fluid-saturated polymer foam scaffold [43] opens up the possibility of probing the internal characteristics of foam tissue scaffolds with acoustics.

**Acknowledgements**
We thank Drs L White and F Rose (Tissue Engineering Group, School of Pharmacy, University of Nottingham) for providing the tissue scaffolds used in this work, and the Biotechnology and Biology Science Research Council for funding (Ref: BB/F004923/1).

**References**

[1]     Langer R and Vacanti J P 1993 *Science* **260** 920-6
[2]     Fisher J P, et al. 2007 *Tissue engineering* (CRC Press, Boca Raton)
[3]     Macchiarini P, et al. 2008 *Lancet* **372** 2023-30
[4]     "UCL surgeons perform revolutionary transplant operation", http://www.ucl.ac.uk/news/news-articles/1003/10031903, accessed 29/03/2010
[5]     Middleton J C and Tipton A J 2000 *Biomaterials* **21** 2335-46
[6]     Ma P X and Elisseeff J H 2005 *Scaffolding in tissue engineering* (Taylor&Francis, Boca Raton)
[7]     Tomlins P, et al. 2004 *Tissue engineered medical products,* ed E Schutte*, et al.* (American Society Testing and Materials, West Conshohocken) pp 3-11
[8]     Mather M L, et al. 2008 *Biomed. Mater.* **3** 11
[9]     Spowage A, et al. 2006 *J. Porous Mat.* **13** 431-8
[10]    Nakao S-i 1994 *J. Membr. Sci.* **96** 131-65
[11]    Piche L 1989 *IEEE Ultrason. Sym*. 1989 600-608
[12]    Bordelius N A, et al. 1973 *Mech. Compo. Mater.* **9** 660-2
[13]    Tatibouet J and Piche L 1991 *Polymer* **32** 3147-51
[14]    Smith D M and Wiggins T A 1972 *Appl. Opt.* **11** 2680-3
[15]    Ding Y, et al. 2004 *Macromolecules* **37** 9264-72
[16]    Bourbie T, et al. 1987 *Acoustics of porous media* (Technip, Paris)
[17]    Haire T J and Langton C M 1999 *Bone* **24** 291-5
[18]    Müllner H W, et al. 2008 *Strain* **44** 153-63
[19]    Richards M S, et al. *2008 IEEE Ultrasonics Symposium* (IEEE, New York) pp 536-9
[20]    Mather M L, et al. 2008 *J. Mater. Sci.-Mater. Med.* **19** 3071-80
[21]    Wu H-C, et al. 2003 *Biomaterials* **24** 3871-6
[22]    Mikos A G, et al. 1994 *Biomaterials* **15** 55-8
[23]    Quirk R A, et al. 2004 *Curr. Opin. Solid State Mat. Sci.* **8** 313-21
[24]    Tai H Y, et al. 2007 *Eur. Cells Mater.* **14** 64-76
[25]    Parker N G, et al. *Meas. Sci. Technol.* **21** 045901
[26]    Bilaniuk N and Wong G S K 1993 *J. Acoust. Soc. Am.* **93** 1609-12
[27]    Bilaniuk N and Wong G S K 1996 *J.Acoust. Soc.Am.* **99** 3257
[28]    Khasanshin T S and Aleksandrov A A 1984 *J. Eng. Phys. Thermophys.* **47** 1046-52
[29]    Rae P J, et al. 2007 *Polymer* **48** 598-615
[30]    Kaye G W C and Laby T H 1995 *Tables of physical and chemical constants* (Longman, New York)
[31]    Mohamed A, et al. 2007 *J. Appl. Polym. Sci.* **106** 1689-96
[32]    Omelczuk M O and McGinity J W 1992 *Pharm Res* **9** 26-32
[33]    Povey M J W, et al. 2005 *Ultrasonics* **43** 219-26
[34]    Park P I P and Jonnalagadda S 2006 *J. Appl. Polym. Sci.* **100** 1983-7
[35]    Paragkumar N T, et al. 2006 *Appl. Surf. Sci.* **253** 2758-64
[36]    Lim L T, et al. 2008 *Prog. Polym. Sci.* **33** 820-52
[37]    Kopinke F D, et al. 1996 *Polym. Degrad.Stabil.* **53** 329-42
[38]    Houchin M L and Topp E M 2009 *J. Appl. Polym. Sci.* **114** 2848-54
[39]    Lu L, et al. 2000 *Biomaterials* **21** 1837-45
[40]    Jiang X-y, et al. 2003 *J. Centr. South Univ. Technol.* **10** 202-6
[41]    Ahmed A R, et al. 2008 *Eur. J. Pharm. Biopharm.* **70** 765-9
[42]    Trask C A and Roland C M 1989 *Macromolecules* **22** 256-61
[43]    Parker N G, Mather M L, Morgan S P and Povey M J W 2010 *J. Physics: Conf. Proc. (submitted)*